\documentclass[letter]{aa}
\usepackage{graphicx}
\usepackage{longtable, lscape}
\usepackage{natbib}

\begin{document}
\title{The globular cluster system of NGC 1316. II - The extraordinary object SH2 
 \thanks{Based
    on observations taken at the European Southern Observatory, Cerro Paranal, Chile,
    under the programmes 082.B-0680, on observations
    taken at the Interamerican Observatory, Cerro Tololo, Chile. Furthermore based on observations made with the NASA/ESA Hubble Space Telescope (PI: A. Sandage, Prop.ID: 7504), and obtained from the Hubble Legacy Archive, which is a collaboration between the Space Telescope Science Institute (STScI/NASA), the Space Telescope European Coordinating Facility (ST-ECF/ESA) and the Canadian Astronomy Data Centre (CADC/NRC/CSA).    }}

\subtitle{}

\author{
T. Richtler     \inst{1} 
\and
B. Kumar \inst{2}
\and 
L. P.  Bassino \inst{3}
\and
 B. Dirsch \inst{4}
 \and
 A. J. Romanowsky \inst{5}
}
\offprints{T. Richtler}

\institute{
Departamento de Astronom\'{\i}a,
Universidad de Concepci\'on,
Concepci\'on, Chile;
tom@astro-udec.cl
\and
Aryabhatta Research Institute of Observational Sciences,
Manora Peak, Nainital - 263129, India 
\and
Facultad de Ciencias Astron\'omicas y Geof\'isicas
de la Universidad Nacional de La Plata; Consejo Nacional de
Investigaciones Cient\'ificas y T\'ecnicas; and Instituto de
Astrof\'isica de La Plata (CCT La Plata-CONICET-UNLP), Argentina
\and
Friedrich-Ebert-Gymnasium 
Bonn, Germany
\and
UCO/Lick Observatory, University of California, Santa Cruz,
CA 95064, USA}

\date{Received  / Accepted }

\abstract
{SH2 has been described as an isolated HII-region, located about 6.5\arcmin\ south of the nucleus of NGC 1316 (Fornax A), a merger remnant in the
the outskirts of the Fornax cluster of galaxies. 
  }
{We give a first, preliminary description of the stellar content and  environment of this remarkable object.
  }
{We used photometric data in the Washington system and  HST photometry from the Hubble Legacy Archive for  a morphological description and preliminary
aperture photometry. Low-resolution spectroscopy provides radial
velocities of the brightest star cluster in SH2 and a nearby intermediate-age cluster. 
    }
{ SH2 is not a normal HII-region, ionized by very young stars. It contains a multitude of  star clusters with ages of approximately $10^8$yr. A
ring-like morphology is striking.   
  SH2 seems to be connected to an intermediate-age massive globular cluster with
a similar radial velocity, which itself is  the main object of a group of fainter clusters. Metallicity estimates from emission lines remain ambiguous.   }
{ The present data do not yet allow firm conclusions about the nature or origin of  SH2.  It  might be a dwarf galaxy that has experienced a burst of extremely clustered star formation. We may witness
 how globular clusters are donated to a parent galaxy.  }

\titlerunning{HII-region SH2}
\maketitle

\section{Introduction}

NGC 1316 (Fornax A) is a merger remnant in the outskirts of the
Fornax cluster. Starting with the classical work of  \citet{schweizer80}, it
has been intensively investigated from X-rays to the radio regime \citep{donofrio95,shaya96,
 mackie98,arnaboldi98,longhetti98,kuntschner00, horellou01,goudfrooij01b,goudfrooij01a,gomez01,
 kim03,bedregal06, nowak08,lanz10, mcneil12}.
 
Morphologically, it is characterized by an inner bulge with 
extensive dust fine structure and a number of shells, ripples and tidal structures at larger
radii. 
 Intermediate-age stellar populations of about 2 Gyr and younger contribute significantly to the
total luminosity (e.g. \citealt{kuntschner00}; \citealt{richtler12}, hereafter Paper I). 

Regarding young populations, one may suspect star formation to be still ongoing in the central
regions, as indicated by the dust and the presence of dense molecular gas \citep{horellou01}.
At larger radii, there is no evidence for really young populations, with one exception: an HII-region
located about 6.5\arcmin\ south of the nucleus, which has been discovered by \citet{schweizer80}
and was labeled SH2 (southern HII-region) by \citet{mackie98}. Its diameter is about 10\arcsec or 860 pc.

Schweizer described the main $H_\alpha$-emission as coming ''..from a banana-shaped..region
along the south-western edge of a blue round fuzz..'' SH2 is one of the few places in NGC 1316 where atomic 
hydrogen has been detected. \citet{horellou01} quote an HI-mass of  $2\times10^7 M_\odot$. The center
of the HI-emission appears to be displaced from the center of SH2 by about 10\arcsec\ to the north-east.
 Molecules have not been detected. Horellou et al. give an upper limit of $7\times10^6 M_\odot$ for
the H$_2$-mass.

In the framework of a  kinematic investigation of the globular cluster system
of NGC 1316, we also obtained spectra for the brightest object in SH2. Moreover, HST/WFPC2 imaging
and photometry in the Washington system is available.
The  aim of this contribution is to give a first  preliminary description of this remarkable object, which may be an example for
a way, how
globular clusters are donated to a parent galaxy. 

We adopt the supernovae Ia distance of 17.8 Mpc  quoted by \citet{stritzinger10} (distance modulus 31.25).
One arcsec corresponds to 86.3 pc.  

\section{Observations}
We found SH2 on archival HST/WFPC2-images, taken on August 8, 1999  (PI: A. Sandage, proposal ID: 7504), at the
image border of exposures in F555W and F814W.
The ground-based photometric and spectroscopic data  are presented in detail
in other contributions (Paper I, Kumar et al. 2012, in prep.),
 therefore we give  some basic information only.
The ground-based photometry uses wide-field imaging in the Washington system with the MOSAIC-camera at the 4m-Blanco telescope
at Cerro Tololo, Chile. The reduction was performed as described in \citet{dirsch03} and Paper I.
 Within a program aiming at measuring radial velocities of globular clusters we obtained spectra  using FORS2/MXU at the VLT. The spectral resolution is 5\AA, using the grism 600B.

\section{Morphology}
One look at Fig.\ref{fig:SH2_image} is sufficient to convince any spectator of the extraordinary nature of this object. 
 We 
see a multitude of  sources, the brightest of them arranged in a ring-like fashion. The brighter objects are marginally resolved. 
At least the brightest object shows a young stellar spectral continuum, therefore we assume that the sources are star clusters and not compact HII-regions
or even extremely bright OB-stars. That these star clusters are young, but not extremely young, is suggested by their colors. The distribution
of point sources is extended toward the south-west, pointing toward another assembly of  sources at a distance of 13\arcsec (1120 pc). This second
group is dominated by a bright globular cluster. See more remarks below. 
We point out that there is no indication that any other object is connected to SH2. The bright source at a distance of 7\arcsec and a position
angle of -35$^\circ$ is a background galaxy at z=0.54.
 
\section{Photometric properties of SH2}
\subsection{HST-imaging and Washington color map}
We name the relevant objects according to our spectral scheme. The brightest source in SH2 is object 431-1 (mask 4, slit 31, and object Nr.1 as the brightest). The two other sources  for which brightnesses  and colors in the Washington system
were measured are indicated by numbers 2 and 3 and the values are given in Table\ref{tab:mags}. 

Owing to the lower resolution of the MOSAIC image, we were unable to measure more sources.
 The magnitudes result from PSF-fitting. It is clear from the comparison with the HST-image that the photometry is
approximate because of the immense crowding, but nevertheless the results are good indications. The colors do  not indicate superbright OB stars or groups of OB
stars, but rather star clusters.  

\begin{figure*}[]
\begin{center}
\includegraphics[width=0.7\textwidth]{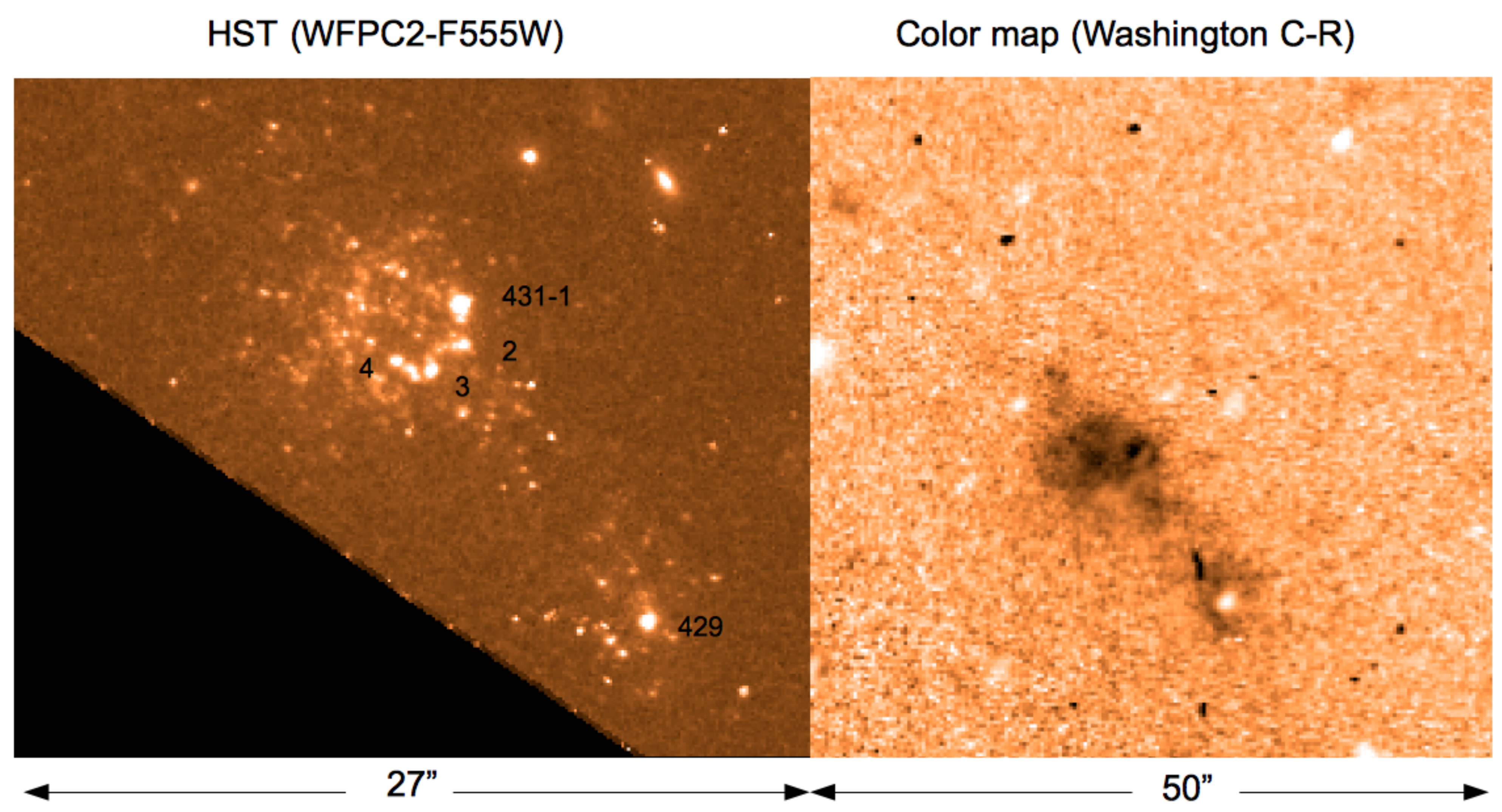}
\caption{Left panel: The region of SH2 as seen on a WFPC2 exposure of 5000 sec (filter F555W). The objects for which photometric values are
given are indicated.  Object 429 is a globular cluster. The 
other objects in the field are either foreground stars or background galaxies. Right panel: Color map in C-R of the same region, constructed from
MOSAIC images with lower spatial resolution. Bright is red, dark is blue with a dynamical range 0 $<$ C-R $<$ 2 mag. The background color is typically C-R=1.3, the darkest spots 
have  C-R=0.2 . The color map is shown with a larger scale  to better demonstrate the color homogeneity of the surrounding and the blueish envelope of SH2.}
\label{fig:SH2_image}
\end{center}
\end{figure*}

The right-hand side  of Fig.\ref{fig:SH2_image} is the color map of the surroundings in  lower resolution (seeing 1.1\arcsec), which we constructed from the C - and R MOSAIC images. The dynamical range is 0 $<$ C-R $<$ 2.  Here we also see  the background stellar population of NGC 1316, which was subtracted
from the HST-image.  Bright is
red, dark is blue. The color of the background stellar population is typically C-R=1.3.
One recognizes that both SH2 and the associated object are surrounded by a region distinctly
bluer than the background with a typical color of C-R=0.9.  The O-II 3727 emission roughly coincides 
with  the maximum transmission of the C-filter, but we cannot easily distinguish between a gaseous
emission and a blue stellar population. However, the HST-image suggests that the bluer colors represent the unresolved 
stellar population.  The color around SH2 is about 0.5 and the two dark
spots show a color of 0.2.  There is no indication for the presence of  dust.

The color of object 429 is surprisingly red and could indicate  a globular cluster  of age 2 Gyr, if  solar metallicity as the typical metallicity
for younger populations in NGC 1316 were assumed (of course, we
cannot exclude an old cluster of intermediate metallicity, in which case it would be an $\omega$-Centauri-like cluster). Object 429 is
embedded in some blue fuzz, which may be an unresolved stellar population, but
a group of sources with colors similar to 429 can also be identified in the CMD (see next section).

 One might doubt a physical connection with SH2, but the morphological indications (the common blueish envelope, the extension of SH2 toward
 429)  
are puzzling. Moreover, the radial velocities are similar (Sec.\ref{sec:spectra}). 

\subsection{Aperture photometry from HST images}
\
We  used the source catalogs extracted with Daophot and SExtractor,  available in the Hubble Legacy Archive.
 This photometry can perhaps be improved by future ACS observations. The objects 429 and 431 are not in the Daophot source list, which 
 otherwise lists many more objects in the most crowded region around 431 than does the SExtractor source list. We therefore used the Daophot
 aperture magnitudes and  
  the SExtractor magnitudes only for 429 and 431. In both cases, the magnitudes were measured through  apertures with radii of 3 pixels.

 The magnitudes may not be
 precise because of   the extreme crowding, but the colors are sufficiently accurate to distinguish
objects that are as different as 431-1 and 429. 

\begin{figure}[]
\begin{center}
\includegraphics[width=0.5\textwidth]{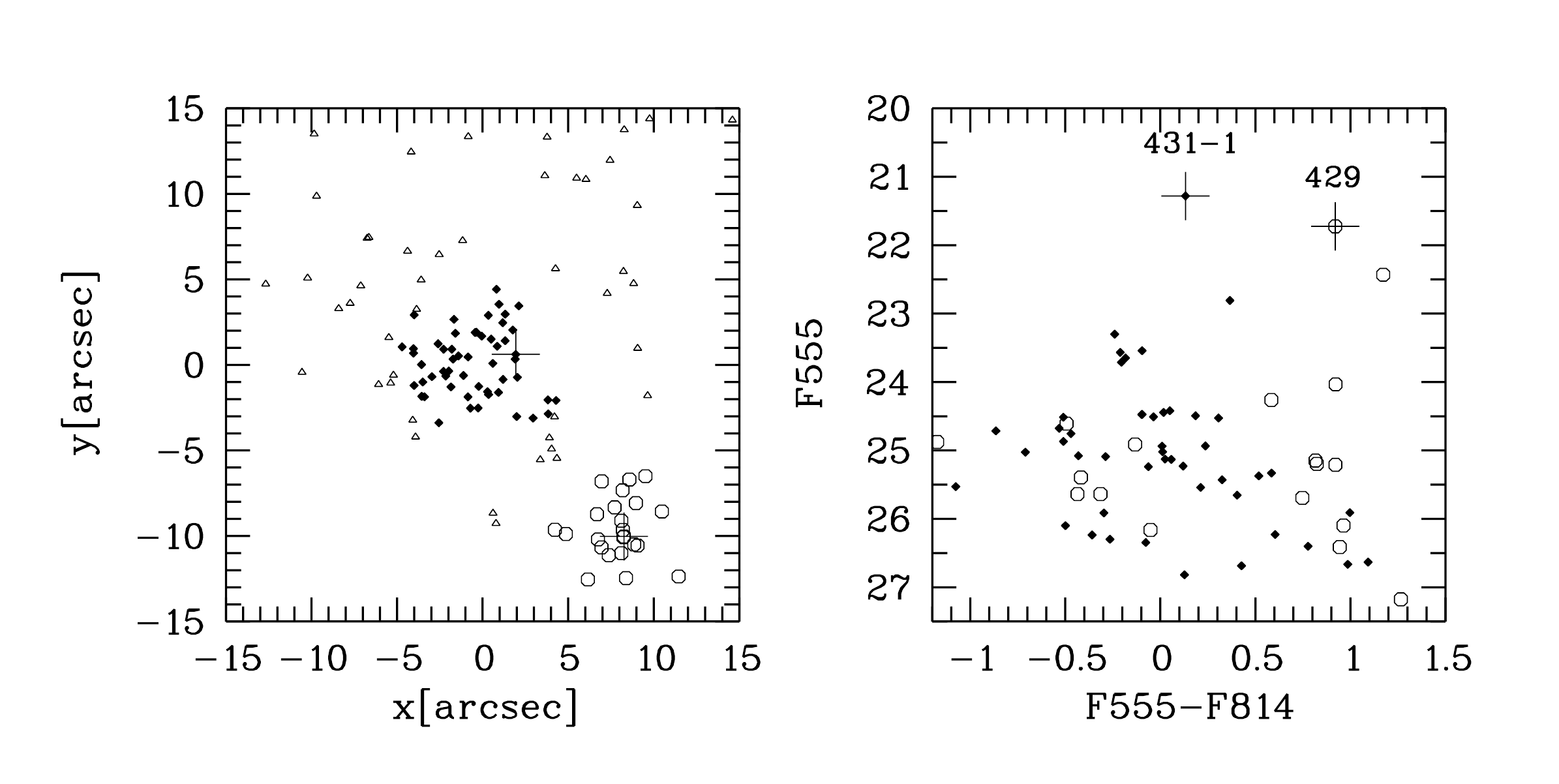}
\caption{Left panel: The locations of objects with available aperture photometry. Symbols are according to their proximity either to 431 (filled circles), 429 (open circles)
or the 'field`` (triangles). Right panel: CMD for the sources from the left panel, using the same symbol coding. The magnitudes are on the ABMAG system and the colors
can  be approximately  transformed into V-I by adding 0.4 mag.
}
\label{fig:HSTCMD}
\end{center}
\end{figure}

The left panel of Fig.\ref{fig:HSTCMD} shows the locations (relative
to an approximate center of SH2) of all 127 objects within a 30\arcsec\ by 30\arcsec\ region, for which photometry in both filters (F555 and F814) are available. 
The sources belonging to SH2 (defined by a radius of 5\arcsec\ )   are denoted by filled circled, the objects belonging to the 429-group by open circles.  The
objects 429 and 431 themselves are crosses. 
The right panel shows the CMD, using the same symbol coding. Magnitudes are given in the ABMAG-system \citep{sirianni05}.

 The color can be approximately transformed into Cousins V-I by adding 0.4 mag.
The large scatter for  fainter objects  may not be real, but produced
by the extreme crowding.
The colors are quite blue, but do not indicate extremely
young objects. It is striking that the faint sources around 429 also fit  to the
redder color of 429 which suggests  a  generically connected ensemble. 
 The nature of these sources is not known, but one may suspect that many of them are fainter GCs somehow related to 429.

\subsection{Stellar masses}
We can roughly estimate stellar masses by comparing magnitudes and colors with theoretical values. Doing that  we assume solar metallicity and
use the C-R color as an age indicator.  A comparison of ground-based and HST photometry  in Table \ref{tab:mags} shows brighter R-magnitudes, 
as expected from the extreme crowding.
Employing the models of \citet{marigo08}, one finds an age of 0.16 Gyr for a color C-R=0.2 (the models use
the  IMF  of \citealt{chabrier01}). The mass-to-light ratio in the R-band is 0.24 (adopting 4.45 as the absolute solar R-magnitude).
We find then  for 431-1 a mass of $1.9 \times 10^5~M_\odot$, using the R-magnitude.   The true brightness and mass  are somewhat   lower. At these blue colors, metallicity is not a critical parameter for the age.

The total absolute R-brightness of SH2 is -12.9, corresponds to about $\mathrm 8\times10^7 L_\odot$. For a single stellar population
and attributing the entire luminosity to stellar emission, one would assign an age of about 0.1 Gyr  and a mass  of  $1.6\times 10^6~M_\odot$.

For 429, one finds an age of 2 Gyr, $M/L_R$=1.1, and a mass of $0.9 \times10^6 M_\odot$ at solar metallicity.
 

\begin{table}[h]
\caption{Washington and HST photometry of sources in SH2}
\begin{center}
\begin{tabular}{ccccc}
Object & mag(R) & C-R & F555 & F555-F814 \\
\hline
429  & 20.83 & 1.52  & 21.73 & 0.92 \\
431-1& 20.88 & 0.20 & 21.28 & 0.13\\
431-2 & 21.71 & 0.25 & 23.56 & -0.21\\
431-3 & 21.19 & 0.37 & 23.30 & -0.24 \\
431-4  &-  &   -  & 23.54 & -0.10 \\
SH2 & 18.4 & 0.06  & - &  - \\
\hline
\end{tabular}
\end{center}
\note{The HST magnitudes are in the ABMAG-system \citep{sirianni05}.The total magnitude and color  of SH2 was measured through an aperture of 12\arcsec\  diameter.}
\label{tab:mags}
\end{table}

\section{Spectral information}
\label{sec:spectra}
\begin{figure}[h]
\begin{center}
\includegraphics[width=0.4\textwidth]{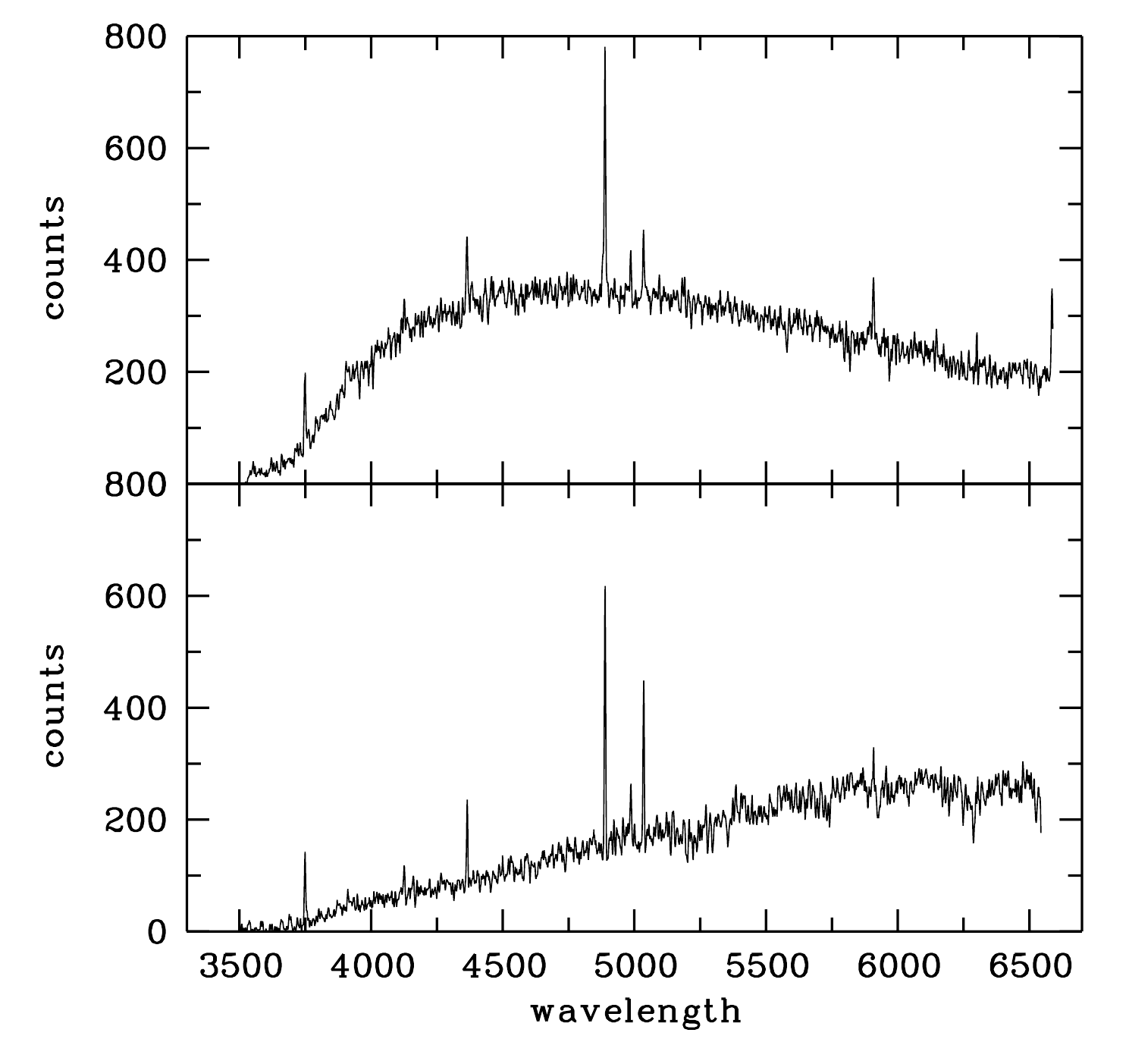}
\caption{Upper panel: spectrum of 431-1. Lower panel: spectrum of 429. The spectra are not flux-calibrated, but the hotter continuum of 431-1
is striking.  }
\label{fig:spectra}
\end{center}
\end{figure}

Slits were centered on 429 and 431-1. The resulting spectra are shown in Fig.
\ref{fig:spectra}. Our spectra are not flux-calibrated but it is clear that the
continuum of 431-1 is much bluer than that of 429, as one expects from the
photometry. This also shows that at least this object is not a compact HII, but a star
cluster. For measuring velocities from emission lines, we used the
following lines:
OII-3727,  HI-4101,HI-4340, HI-4861, OIII-4959, OIII-5007, HeI-5876. The
resulting heliocentric velocities are listed in Table \ref{tab:velocities}.
Strikingly, the two bluest lines (3727 and 4101) are offset from the others,
which agree well. Within the wavelength calibration,
the bluest line was 3888, so some extrapolation is necessary to include
3727. It is not clear why H$_\delta$ deviates.
Skipping OII-3727 and  H$_\delta$, we obtain  1709 $\pm$ 12 km/s as the mean velocity of the
emission lines in 429 and  1660$\pm$11  km/s as the corresponding velocity for 431.
 
To measure the radial velocity of the stellar absorption spectrum for 429, we removed the emission
lines in the region 4800-5400 for 429 and cross-correlated this spectral region
with a template spectrum, obtained with the same instrumentation that was already
used in \citet{schuberth10}.

We measure a velocity of 1685$\pm$ 20 km/s. The difference of 24 km/s to the line velocities
is hardly significant. Certainty will only come from spectroscopy of higher spectral resolution.

The color of 431-1 indicates a spectral type  of about early A. The strong hydrogen absorption lines which we would
expect are filled up by emissions. CaII H+K as well as molecular bands (G-band at 4300 \AA\ and the CN-band at 3883 \AA) are weak,
but visible.
 One can conclude at least that
the spectrum is not of O-type or early B-type and that it shows  a  velocity very similar to that of the emitting gas.
Taking the mean of CaII H+K, we obtain a radial velocity of 1670 $\pm$20 km/s.

\citet{schweizer80} quotes a velocity of SH2 relative to the core of NGC 1316 of -101$\pm$8 km/s.
That leads to 1760 km/s for the radial velocity of NGC 1316 and agrees well with all previous measurements.
The HI-velocity of SH2 is  1690 km/s \citep{horellou01} and fits excellently.

\begin{table}[h]
\caption{Velocities of spectral lines and equivalent widths for 431-1 and 429. The equivalent widths are rough estimates only.}
\begin{center}
\begin{tabular}{cccccc}
Line & rest wavelength & 429 & 431-1 & EW(429) & EW(431)\\ 
\hline
  OII &         3727.32& 1745 &1737 &   37 & 18.5\\
  H$_\delta$ &      4101.73& 1779 &1769 &  - & -\\
H$_\gamma$ &4340.47 & 1721 &1645 & - & -\\
H$_\beta$ &4861.33&  1698 &1674 & 16 & 8.1 \\
   OIII    &  4958.91&  1716 & 1665 & 2.8 & 2.0 \\
    OIII&     5006.84&  1717 &1663 & 9.5 & 2.3 \\
   HeI &      5875.69&  1691 &1651 & - & - \\
\hline
\end{tabular}
\end{center}
\label{tab:velocities}
\end{table}

The metallicity of SH2  obviously is a key parameter: For a dwarf galaxy, one would expect an abundance distinctly lower  than solar. As \citet{kobulnicky03} showed, simple  diagnostics based on the equivalent widths of $H_\beta$ and the [OIII], [OII]-lines (the $\mathrm  R_{23}$-parameter) lead to good results for
HII regions. Unfortunately, our spectra miss
the quality for such work and the quoted equivalent widths in Table \ref{tab:velocities} are only rough estimates. Moreover, the continuum contamination 
of H$_\beta$ in 431 is unknown. However, it is striking that the OIII-line 5007 is much weaker with respect to $H_\beta$ than in typical HII-regions. 
The R$_{23}$-parameter is defined as $\mathrm R_{23} = log  \left ( (EW(3727)+EW(4958) + EW(5007))/EW(H_\beta)) \right )$. A calibration in terms
of oxygen abundance is given by \citet{kobulnicky99} (their Fig.8). The $\mathrm  R_{23}$-values for 431-1 and 429 are 0.24 and 0.49, respectively.
They indicate either a super-solar or a very low metallicity due to the ambiguity of  $\mathrm  R_{23}$ without other diagnostic parameters. 

\section{Discussion and conclusions}
This is  a contribution where questions are mainly posed, not answered.  The morphology of SH2 gives the impression of a region of recent star-formation
 (the only one outside the nucleus of NGC 1316) with extreme clustering.  The  colors  of the star  clusters indicate an age of about 0.1 Gyr, thus
SH2 is not a typical  HII-region, where star formation is ongoing and the ionization sources
are UV-photons from very young stars.   
Very puzzling is its apparent association with a star cluster  that is much older. This is hard to understand, even if 
one speculates that SH2 is the debris of a dwarf galaxy and the cluster 429 one of its globular clusters, which
dynamically survived the disruption process.  Such a massive object should not come without a field population. 

On the other hand,  the similarity of the radial velocities can be understood if both
SH2 and 429 are moving within a disk  with only a low velocity dispersion along the
line of sight. But if on excludes  429 itself as  the ionizing  source,
what ionizes the gas around 429? 

The grouping of fainter sources with similar colors around 429 is another puzzling finding.  




What is the dynamical state and fate of SH2? If it disperses, it may donate a significant number of star clusters to the
cluster system of NGC 1316. Why the ring-like structure? How high  is the metallicity? What ionizes the gas? Is the connection
with 429 real?  Is there a faint older population? These questions may find answers after a thorough investigation with integral field spectroscopy and
high-resolution space-based and ground-based adaptive optics imaging.

\begin{acknowledgements}
We cordially thank the referee, Sidney van den Bergh, for his report. TR acknowledges financial support from the Chilean Center for Astrophysics,
FONDAP Nr. 15010003,  from FONDECYT project Nr. 1100620, and
from the BASAL Centro de Astrofisica y Tecnologias
Afines (CATA) PFB-06/2007. He also thanks the Aryabhatta Research Institute for Observational Sciences, Nainital, for warm hospitality 
and financial support.  LPB gratefully acknowledges support
by grants from Consejo Nacional de Investigaciones Cient\'ificas
y T\'ecnicas and Universidad Nacional de La Plata (Argentina).
We thank Fabio Bresolin for advice on spectral diagnostics of HII-regions and Thomas Puzia for remarks on a draft version.
\end{acknowledgements}
\bibliographystyle{aa}
\bibliography{N1316.bib}

\begin{thebibliography}{25}
\expandafter\ifx\csname natexlab\endcsname\relax\def\natexlab#1{#1}\fi

\bibitem[{{Arnaboldi} {et~al.}(1998){Arnaboldi}, {Freeman}, {Gerhard},
  {Matthias}, {Kudritzki}, {M{\'e}ndez}, {Capaccioli}, \& {Ford}}]{arnaboldi98}
{Arnaboldi}, M., {Freeman}, K.~C., {Gerhard}, O., {et~al.} 1998, \apj, 507, 759

\bibitem[{{Bedregal} {et~al.}(2006){Bedregal}, {Arag{\'o}n-Salamanca},
  {Merrifield}, \& {Milvang-Jensen}}]{bedregal06}
{Bedregal}, A.~G., {Arag{\'o}n-Salamanca}, A., {Merrifield}, M.~R., \&
  {Milvang-Jensen}, B. 2006, \mnras, 371, 1912

\bibitem[{{Chabrier}(2001)}]{chabrier01}
{Chabrier}, G. 2001, \apj, 554, 1274

\bibitem[{{Dirsch} {et~al.}(2003){Dirsch}, {Richtler}, {Geisler}, {Forte},
  {Bassino}, \& {Gieren}}]{dirsch03}
{Dirsch}, B., {Richtler}, T., {Geisler}, D., {et~al.} 2003, \aj, 125, 1908

\bibitem[{{D'Onofrio} {et~al.}(1995){D'Onofrio}, {Zaggia}, {Longo}, {Caon}, \&
  {Capaccioli}}]{donofrio95}
{D'Onofrio}, M., {Zaggia}, S.~R., {Longo}, G., {Caon}, N., \& {Capaccioli}, M.
  1995, \aap, 296, 319

\bibitem[{{G{\'o}mez} {et~al.}(2001){G{\'o}mez}, {Richtler}, {Infante}, \&
  {Drenkhahn}}]{gomez01}
{G{\'o}mez}, M., {Richtler}, T., {Infante}, L., \& {Drenkhahn}, G. 2001, \aap,
  371, 875

\bibitem[{{Goudfrooij} {et~al.}(2001{\natexlab{a}}){Goudfrooij}, {Alonso},
  {Maraston}, \& {Minniti}}]{goudfrooij01a}
{Goudfrooij}, P., {Alonso}, M.~V., {Maraston}, C., \& {Minniti}, D.
  2001{\natexlab{a}}, \mnras, 328, 237

\bibitem[{{Goudfrooij} {et~al.}(2001{\natexlab{b}}){Goudfrooij}, {Mack},
  {Kissler-Patig}, {Meylan}, \& {Minniti}}]{goudfrooij01b}
{Goudfrooij}, P., {Mack}, J., {Kissler-Patig}, M., {Meylan}, G., \& {Minniti},
  D. 2001{\natexlab{b}}, \mnras, 322, 643

\bibitem[{{Horellou} {et~al.}(2001){Horellou}, {Black}, {van Gorkom}, {Combes},
  {van der Hulst}, \& {Charmandaris}}]{horellou01}
{Horellou}, C., {Black}, J.~H., {van Gorkom}, J.~H., {et~al.} 2001, \aap, 376,
  837

\bibitem[{{Kim} \& {Fabbiano}(2003)}]{kim03}
{Kim}, D.-W. \& {Fabbiano}, G. 2003, \apj, 586, 826

\bibitem[{{Kobulnicky} {et~al.}(1999){Kobulnicky}, {Kennicutt}, \&
  {Pizagno}}]{kobulnicky99}
{Kobulnicky}, H.~A., {Kennicutt}, Jr., R.~C., \& {Pizagno}, J.~L. 1999, \apj,
  514, 544

\bibitem[{{Kobulnicky} \& {Phillips}(2003)}]{kobulnicky03}
{Kobulnicky}, H.~A. \& {Phillips}, A.~C. 2003, \apj, 599, 1031

\bibitem[{{Kuntschner}(2000)}]{kuntschner00}
{Kuntschner}, H. 2000, \mnras, 315, 184

\bibitem[{{Lanz} {et~al.}(2010){Lanz}, {Jones}, {Forman}, {Ashby}, {Kraft}, \&
  {Hickox}}]{lanz10}
{Lanz}, L., {Jones}, C., {Forman}, W.~R., {et~al.} 2010, \apj, 721, 1702

\bibitem[{{Longhetti} {et~al.}(1998){Longhetti}, {Rampazzo}, {Bressan}, \&
  {Chiosi}}]{longhetti98}
{Longhetti}, M., {Rampazzo}, R., {Bressan}, A., \& {Chiosi}, C. 1998, \aaps,
  130, 267

\bibitem[{{Mackie} \& {Fabbiano}(1998)}]{mackie98}
{Mackie}, G. \& {Fabbiano}, G. 1998, \aj, 115, 514

\bibitem[{{Marigo} {et~al.}(2008){Marigo}, {Girardi}, {Bressan}, {Groenewegen},
  {Silva}, \& {Granato}}]{marigo08}
{Marigo}, P., {Girardi}, L., {Bressan}, A., {et~al.} 2008, \aap, 482, 883

\bibitem[{{McNeil-Moylan} {et~al.}(2012){McNeil-Moylan}, {Freeman},
  {Arnaboldi}, \& {Gerhard}}]{mcneil12}
{McNeil-Moylan}, E.~K., {Freeman}, K.~C., {Arnaboldi}, M., \& {Gerhard}, O.~E.
  2012, ArXiv e-prints

\bibitem[{{Nowak} {et~al.}(2008){Nowak}, {Saglia}, {Thomas}, {Bender},
  {Davies}, \& {Gebhardt}}]{nowak08}
{Nowak}, N., {Saglia}, R.~P., {Thomas}, J., {et~al.} 2008, \mnras, 391, 1629

\bibitem[{{Richtler} {et~al.}(2012){Richtler}, {Bassino}, {Dirsch}, \&
  {Kumar}}]{richtler12}
{Richtler}, T., {Bassino}, L.~P., {Dirsch}, B., \& {Kumar}, B. 2012, \aap,
  accepted, Paper I

\bibitem[{{Schuberth} {et~al.}(2010){Schuberth}, {Richtler}, {Hilker},
  {Dirsch}, {Bassino}, {Romanowsky}, \& {Infante}}]{schuberth10}
{Schuberth}, Y., {Richtler}, T., {Hilker}, M., {et~al.} 2010, \aap, 513, A52+

\bibitem[{{Schweizer}(1980)}]{schweizer80}
{Schweizer}, F. 1980, \apj, 237, 303

\bibitem[{{Shaya} {et~al.}(1996){Shaya}, {Dowling}, {Currie}, {Faber}, {Ajhar},
  {Lauer}, {Groth}, {Grillmair}, {Lynd}, \& {O'Neil}}]{shaya96}
{Shaya}, E.~J., {Dowling}, D.~M., {Currie}, D.~G., {et~al.} 1996, \aj, 111,
  2212

\bibitem[{{Sirianni} {et~al.}(2005){Sirianni}, {Jee}, {Ben{\'{\i}}tez},
  {Blakeslee}, {Martel}, {Meurer}, {Clampin}, {De Marchi}, {Ford}, {Gilliland},
  {Hartig}, {Illingworth}, {Mack}, \& {McCann}}]{sirianni05}
{Sirianni}, M., {Jee}, M.~J., {Ben{\'{\i}}tez}, N., {et~al.} 2005, \pasp, 117,
  1049

\bibitem[{{Stritzinger} {et~al.}(2010){Stritzinger}, {Burns}, {Phillips},
  {Folatelli}, {Krisciunas}, {Kattner}, {Persson}, {Boldt}, {Campillay},
  {Contreras}, {Krzeminski}, {Morrell}, {Salgado}, {Freedman}, {Hamuy},
  {Madore}, {Roth}, \& {Suntzeff}}]{stritzinger10}
{Stritzinger}, M., {Burns}, C.~R., {Phillips}, M.~M., {et~al.} 2010, \aj, 140,
  2036

\end{thebibliography}

\end{document}